\documentclass[amsmath, amssymb, superscriptaddress, reprint, twocolumn, colorlinks=true, citecolor=blue, linkcolor=blue, urlcolor=blue]{revtex4-1}

\usepackage{mathtools}
\usepackage{natbib}
\usepackage[utf8]{inputenc}
\usepackage[T1]{fontenc}
\usepackage{graphicx}
\usepackage{grffile}
\usepackage{longtable}
\usepackage{wrapfig}
\usepackage{rotating}
\usepackage[normalem]{ulem}
\usepackage{amsmath}
\usepackage{textcomp}
\usepackage{amssymb}
\usepackage{capt-of}
\usepackage{hyperref}
\usepackage{mathptmx}
\usepackage{upgreek}
\usepackage{float}
\usepackage{bm}
\usepackage{setspace}
\usepackage{booktabs}
\usepackage{subfigure}
\usepackage{url}
\usepackage{xcolor}
\usepackage{multirow}
\usepackage{array}
\usepackage{tabularx}
\usepackage{nicematrix}
\usepackage{blkarray}
\usepackage{mathptmx}
\usepackage{mathrsfs}
\usepackage{xcolor}
\makeatletter
\newcommand{\coloredcite}[1]{{\color{ccr}\cite{#1}}}
\newcommand{\coloredeqref}[1]{{\color{ccr}\eqref{#1}}}
\makeatother

\definecolor{ccr}{RGB}{45,47,145} 

\hypersetup{colorlinks=true, citecolor=ccr, urlcolor=ccr, linkcolor=ccr}

\DeclareMathAlphabet{\mathcal}{OMS}{cmsy}{m}{n}

\date{\today}

\begin{document}

\title{Similarity Theory and Scaling Networks for Electromagnetic Wave-Driven Plasmas}

\author{Hanyang Li}
\affiliation{Department of Electrical Engineering, Tsinghua University, Beijing 100084, China}

\author{Yulia Sharova}
\affiliation{Institute of Theoretical Electrical Engineering, Ruhr University Bochum, Universitätsstrasse 150, D-44801 Bochum, Germany}

\author{Denis Eremin}
\email{Denis.Eremin@rub.de}
\affiliation{Institute of Theoretical Electrical Engineering, Ruhr University Bochum, Universitätsstrasse 150, D-44801 Bochum, Germany}

\author{Yangyang Fu}
\email{fuyangyang@tsinghua.edu.cn}
\affiliation{Department of Electrical Engineering, Tsinghua University, Beijing 100084, China}
\affiliation{State Key Laboratory of Power System Operation and Control, Department of Electrical Engineering, Tsinghua University, Beijing 100084, China}
\affiliation{Sichuan Energy Internet Research Institute, Tsinghua University, Sichuan 610213, China}

\begin{abstract}


We demonstrate the scale-invariant behavior of electromagnetic wave–driven radio-frequency plasmas across different dimensional scales. Using two-dimensional electromagnetic particle-in-cell simulations, we show that plasma uniformity remains the same in similar discharges. Building on the concept of similarity laws, we develop scaling networks that effectively relate plasma parameters across varying operating conditions. These results establish a generalized similarity theory derived from the Boltzmann equation coupled with the full set of Maxwell equations, extending the theoretical framework of similarity laws into electromagnetic regimes.

\end{abstract}

\maketitle
\textit{Introduction}---
Similarity theory is based on the principle that physical laws remain unchanged under appropriately scaled conditions \coloredcite{Bluman,Buckingham}, providing a powerful framework for predicting, comparing, and correlating discharges across systems of different sizes. The seminal studies of similarities in gas discharge can be found in Townsend \coloredcite{Townsend-596}, 
Francis \coloredcite{Francis}, 
von Engel \coloredcite{Engel-597}, and many others \coloredcite{GordonWhite-598,Muehe-597,MichaelKhodorkovskii-600,Ryutov2018}. 
Similarity properties in discharge physics have been demonstrated in glow discharges \coloredcite{Gudmundsson_2017, Janasek-nature}, streamers \coloredcite{LiuPasko, Zhao-TDEI}, pulsed discharges \coloredcite{Mesyats-2006}, and high-frequency discharges \coloredcite{Margenau-706}.
For radio-frequency (rf) discharges, Lisovskiy \textit{et al.} \coloredcite{Lisovskiy_2008EPL} experimentally demonstrated that the breakdown scaling curves overlap when the combined parameters \(pd\) (pressure times dimension) and \(fd\) (frequency times dimension) are kept the same in compared gaps.
However, conventional similarity laws are established based on the assumptions of local-field or local energy approximations.
Previously, we demonstrated the similarity laws for rf plasmas with highly nonlocal electron kinetics via particle-in-cell (PIC) simulations  \coloredcite{FuZheng-pop2020}, and later extended the similarity scaling for rf plasma across nonlinear transitional regimes, such as the alpha-gamma mode transition, the stochastic-Ohmic-heating mode transition, and the bounce-resonance-heating mode transition \coloredcite{fu-prapplied}.
Following that, the similarity laws have been applied to inductive \coloredcite{ICP01,Wen_2025} and capacitive rf discharges, considering the electrical and geometrical asymmetry effects \coloredcite{Yang_2022,Zhang-JAP}, and nonlinear collision effects \cite{Yd-psst}. 

Most recently, some of the authors have experimentally demonstrated similarity laws for capacitive rf plasmas using the phase-resolved optical spectroscopy, which validated electrostatic PIC simulations \coloredcite{PRL}. 
However, with the growing interest in large-area uniform plasma sources, electromagnetic effects (e.g., standing wave effects \coloredcite{Rauf_2008,Lee_2008,zhao-PhysRevLett,Eremin_2023}) are found to be prominent in high-frequency discharges since the electrode radius becomes comparable to the wavelength of surface modes existing due to the bulk plasma being separated from electrodes by electron-depleted sheaths. This can strongly deteriorate the plasma uniformity critical for plasma processing \coloredcite{JVST2013,JVST2024}.
To date, the similarity laws for plasmas have not yet been demonstrated in the electromagnetic regimes.


In this Letter, we present the similarity theory and corresponding scaling networks for electromagnetic wave–driven plasmas. Using electromagnetic PIC simulations, we demonstrate scale invariance of electron density and electric field across geometrically similar discharges. We further develop  
similarity-based scaling networks for the plasma uniformity coefficient, enabling effective correlation of plasma characteristics under different operating conditions. Consequently, this work generalizes the similarity framework from electrostatic to electromagnetic regimes and provides new scaling strategies for high-frequency plasma systems.

\textit{Theory}---The Boltzmann equation \coloredcite{bitten} for describing the distribution functions of species in plasmas is given by
\begin{equation}
\frac{\partial f}{\partial t} + \mathbf{v} \cdot \frac{\partial f}{\partial \mathbf{r}} + \frac{q}{m}(\mathbf{E}+\mathbf{v}\times \mathbf{B}) \cdot \frac{\partial f}{\partial \mathbf{v}}  = \left(\frac{\partial f}{\partial t}\right)_{\text{coll}},
\label{eq:boltzmann0}
\end{equation}
where \(f=f_j\) with \(j\in \{e, i\}\) denotes the distribution function for electron or ion species; \(\mathbf{v}\) is the velocity, \( q\) is the particle charge, and \(m\) is the particle mass; \(\mathbf{E}\) and \(\mathbf{B}\) are the electric and magnetic fields.
Taking the electron for example, the collisional term on the right-hand side of Eq.~\coloredeqref{eq:boltzmann0} is expressed as
\begin{equation}
\left(\frac{\partial f_{e}}{\partial t}\right)_{\text{coll}}=\sum\iint\left(f_{e2}f_{n2}-f_{e1}f_{n1}\right)\mathbf{v}_\text{r}\sigma\text{d}\Omega\text{d}\mathbf{v},
\label{eq:collision0}
\end{equation}
where \( f_{{n1}}\) and \(f_{{n2}} \) are the neutral particle distribution functions before and after collisions, respectively; \( f_{e1}\) and \(f_{e2} \) denote the distribution functions of electrons before and after collisions, respectively; \(\mathbf{v}_\text{r}\) is the electron-neutral relative velocity, \(\sigma=\sigma(\mathbf{v}_\text{r},\Omega) \) is
 the differential cross section, and \( \text{d}\Omega\) is the solid angle element.
Substituting Eq.~\coloredeqref{eq:collision0} into Eq.~\coloredeqref{eq:boltzmann0} and dividing both sides by \(k^3\) gives
\begin{align}
&\frac{\partial (k^{-2}f_{e})}{\partial (kt)} + \mathbf{v} \cdot \frac {\partial (k^{-2}f_{e})}{\partial (k\mathbf{r})}  + \frac{e}{m_e}\left({\frac{\mathbf{E}}{k}+\mathbf{v}\times\frac{\mathbf{B}}{k}}\right)\cdot   \frac{\partial (k^{-2}f_{e})}{\partial \mathbf{v}} \nonumber\\ 
&= \sum\iint\left[(k^{-2}f_{e2})\frac{f_{n2}}{k}-(k^{-2}f_{e1})\frac{f_{n1}}{k}\right]\mathbf{v}_\text{r}\sigma\text{d}\Omega\text{d}\mathbf{v},
\label{eq:boltzmann02}
\end{align}
where \(k\) is the scaling factor, independent of the variables in the Boltzmann equation.
The scaled variables, \(k^{-2}f_e\), \(kt\), \(k\mathbf{r}\), \(k^{-1}\mathbf{E}\), \(k^{-1}\mathbf{B}\), and \(k^{-1}f_n\), enable the proportional mathematical solutions of the Boltzmann equation when the external fields and geometrical dimensions are simultaneously scaled.

In the following, we demonstrate that the scaling of the Boltzmann equation is consistent when coupling to Maxwell's equations, thereby extending the scaling formalism to electromagnetic fields.
The Maxwell's equations can be scaled by dividing \(k^2\) and formulated as
\begin{subequations}
\label{eq:Maxwell}
\begin{align}
 \nabla_{k\mathbf{r}}\cdot (k^{-1}\mathbf{E}) 
  & =\frac{1}{\varepsilon_0}\sum_{j}q_j\int k^{-2}f_{j} \mathrm{d}\mathbf{v},\\
 \nabla_{k\mathbf{r}}\cdot(k^{-1}\mathbf{B}) 
&  = 0 ,\\
 \nabla_{k\mathbf{r}}\times(k^{-1}\mathbf{E}) 
 & = \frac{\partial (k^{-1}\mathbf{B})}{\partial (kt)},\\
 \nabla_{k\mathbf{r}}\times (k^{-1}\mathbf{B}) 
 & ={\mu_0}\sum_{j }^{} q_j \int  k^{-2} f_j\mathbf{v}\text{d}\mathbf{v} +{\mu_0}\varepsilon_0\frac{\partial (k^{-1}\mathbf{E})}{\partial (kt)},
\end{align}
\end{subequations}
where \(j \in \{e,i\}\), \(\varepsilon_0\) is the vacuum permittivity and \(\mu_0\) is the vacuum permeability.
From Eqs.~\coloredeqref{eq:boltzmann02}--\coloredeqref{eq:Maxwell}, the variables scaled with \(k\)-factors, such as \(k^{-2}f_e\), \(kt\), \(k\mathbf{r}\), \(k^{-1}\mathbf{E}\), \(k^{-1}\mathbf{B}\), and \(k^{-1}f_n\), ensure the same mathematical solution of the Boltzmann-Maxwell equations as their original counterparts. 

The principle of scale invariance in the Boltzmann-Maxwell equations establishes similarity transformations, which indicate that plasma parameters remain dynamically similar in systems of different sizes.
Mathematically, the physical parameter \(G(\mathbf{r},t)\) at the corresponding spatiotemporal points, \((\mathbf{r}_1, t_1) = k(\mathbf{r}_k, t_k)\), can be transformed by
\begin{equation}
G(\mathbf{r}_1, t_1) = k^{\alpha[G]} G(\mathbf{r}_k, t_k),
\label{eq-G1Gk}
\end{equation}
where the subscripts 1 and \(k\) represent the prototype and downscaled (or upscaled) systems, respectively, and \(\alpha[G]\) is the similarity factor for parameter \(G\). 
From Eqs.~\eqref{eq:boltzmann02}--\eqref{eq:Maxwell}, one can determine \(\alpha[\mathbf{r}]=\alpha[t] = 1 \), \(\alpha[\mathbf{v}] = 0 \) (similarity invariants \coloredcite{fu-RMPP2023}), \(\alpha[f_n] = \alpha[\mathbf{E}] = \alpha[\mathbf{B}] = -1\), and \(\alpha[f_e] = \alpha[{f}_i] = -2\) \coloredcite{fu-RMPP2023,Yang-2023,FuZheng-711}.
Then we have \(\alpha[p]=-1\) for pressure (\(p\varpropto f_n\)),  \(\alpha[d]=\alpha[\mathbf{r}]=1\) for dimension, \(\alpha[f]=-\alpha[1/t] = -1\) for frequency, and \(\alpha[n_e]=\alpha[n_i]=-2\) for electron and ion density (\(n_e \varpropto f_e\), \(n_i \varpropto f_i\)).
From Eq.~\coloredeqref{eq-G1Gk}, the combined parameters \(pd\), \(\mathbf{E}/p\) and \(f/p\) can be determined as similarity invariants, i.e., \(\alpha[pd]=\alpha[p]+\alpha[d]=0\) for reduced length, \(\alpha[\mathbf{E}/p]=\alpha[\mathbf{E}]-\alpha[p]=0\) for reduced electric field and \(\alpha[f/p]=\alpha[f]-\alpha[p]=0\) for reduced frequency.
From Eqs.~\eqref{eq:boltzmann02}--\eqref{eq:Maxwell}, the discharge characteristics in two rf systems with the same \(pd\) and \(f/p\) (or \(fd\)) are expected to be scale-invariant. In the following, we demonstrate that the similarity laws for rf discharges hold in the electromagnetic regime, especially when the standing wave effects become prominent.

\begin{figure}[htbp]
\hypertarget{fig1}{}
\centering
\includegraphics[clip, width=0.85\linewidth]{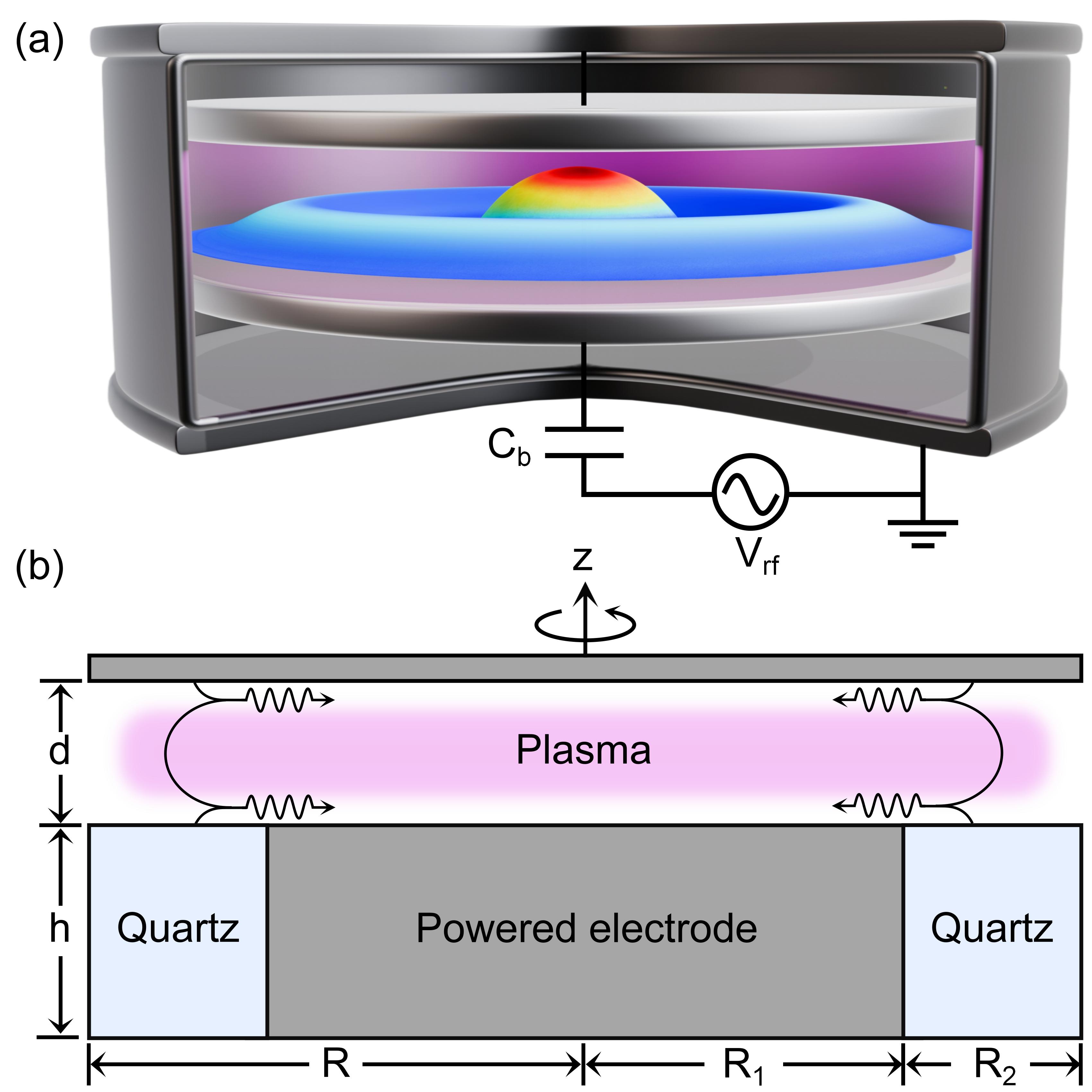}
\caption{\label{fig:Fig1}(a) Schematic of radially nonuniform rf plasma density with electromagnetic wave effects. (b)  Dimensions of the 2D computational domain for the electromagnetic wave-driven rf plasmas.}
\end{figure}

\textit{Model}---Figure~\hyperlink{fig1}{1(a)} shows the schematic of a capacitive rf plasma with radial non-uniformity in the electromagnetic regime. The upper electrode is grounded, whereas the bottom electrode is connected with rf power \(V_{\text{rf}}\) through a blocking capacitor \(C_{\text{b}}\). The color distribution in the figure emphasizes the radial non-uniformity of the plasma, with distinct features in plasma parameters between the central and edge regions. Figure~\hyperlink{fig1}{1(b)} illustrates the simulation model for the rf plasma driven by the electromagnetic wave. 
The radial dimension is \(R\) with the bottom electrode radius \(R_1\) and the quartz radial dimension \(R_2\), i.e., \(R=R_1+R_2\).
The inter-electrode distance is \(d\), and the height of the bottom electrode is \(h\).
Here, the plasma is directly driven by the electromagnetic wave in the \(\text{TEM}_{00}\) mode, instead of using the rf
voltage source.
The electromagnetic wave is deposited via the quartz ports, and the total power absorbed from the wave by plasma is kept at 100 W for all the studied cases.

The two-dimensional (2D) electromagnetic PIC code, ECCOPIC2M (see Refs.~\coloredcite{Eremin_2023, Eremin2025, Eremin2025a} for details on the code and its validation), is employed for simulating the electromagnetic wave-driven rf plasmas. 
In the model, argon at 300 K is adopted as the working gas. The electron-neutral collisions (elastic, excitation, and ionization) and ion-neutral collisions (isotropic and backward scattering) are implemented \coloredcite{Eremin_2023}. 
The grid numbers are \(324\times258\) (\(r \times z\)) in the 2D plasma computational domain.
The secondary electron emission is not expected to play a large role and is omitted.

To achieve similar discharges, two geometrically similar systems are designed for comparison.
The control parameters of rf plasmas \(C_{000}=[p, g, f]\) in the base case, where \(g=(d, R_1, R_2, h)\) denotes the geometrical dimensions, are simultaneously tuned as \(C_{111}=[kp, k^{-1}g, kf]\). Here the scaling factor \(k\), as shown in Eq.~\eqref{eq:boltzmann02}, is 2.
In the base case, \([p, g, f]=[40\ \text{mTorr}, (6\ \text{cm}, 28\ \text{cm}, 10\ \text{cm}, 14\ \text{cm}), 106 \ \text{MHz}]\) while in the down-scaled similarity case, \([p, g, f]=[80 \ \text{mTorr}, (3 \ \text{cm}, 14\ \text{cm}, 5 \ \text{cm}, 7\ \text{cm}), 212\ \text{MHz}]\). 
Since \([p, g, f]\) can also be independently tuned, one can use a binary digit (0 or 1) to represent whether the control parameter is varied. Here, we label the base case as (000) and the similarity case as (111) (also see the control parameter array \(C_{000}\) and \(C_{111}\)). 
In addition, six intermediate cases, i.e., (100), (010), (001), (011), (101), and (110) with only one or two parameters tuned, are studied to fully establish the similarity-based scaling networks.

\begin{figure}[htbp]
\hypertarget{fig2}{}
\centering
\includegraphics[clip, width=\linewidth]{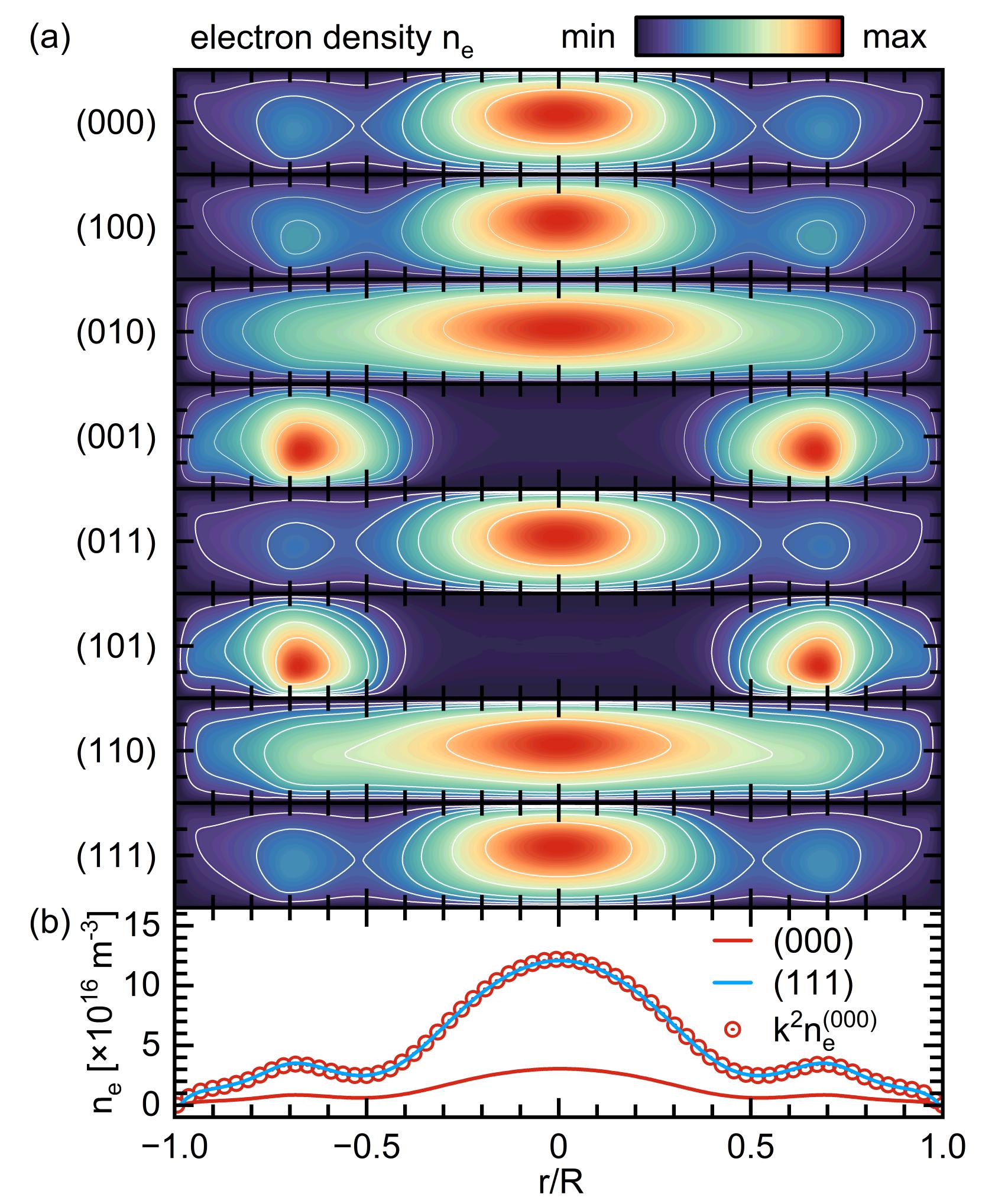}
\caption{\label{fig:Fig2}
(a) Distributions of normalized electron density for the base case (000), the six intermediate cases (100), (010), (001), (011), (101), (110), and the similarity case (000). (b) Radial distributions of the electron density at \(z=d/2\).
The scaled electron density distributions for the base case and similarity case overlaps, which confirms the scale-invariant nature in similar rf discharges.}
\end{figure}

\textit{Similarity and scaling networks}---The distributions of electron density across different discharge conditions are shown in Fig.~\hyperlink{fig2}{2}.
A distinct electromagnetic wave-induced plasma nonuniformity can be observed, for example, in case (000), exhibiting multiple density peaks radially in the plasma bulk.
The electron density, though the highest at the gap center, decreases radially from the center, reaching a local minimum at \(r/R=0.52\), and then increases, forming the second peak at \(r/R=0.69\). 
This plasma nonuniformity is due to the radial standing wave effect when the reactor size approaches the excitation wavelength of the electromagnetic surface waves.
According to the transmission line model \coloredcite{Chabert-2004}, the surface mode excitation wavelength can be determined by \(\lambda_\text{exc} \approx 40d^{-1/2}f^{-2/5}V_0^{1/10}\lambda_0\), where \(d\) is the gap distance, \(f\) is the driving frequency, \(V_0\) is the voltage amplitude, and \(\lambda_0\) is the vacuum wavelength of the electromagnetic wave.
Under the discharge condition of case (000), the excitation wavelength \(\lambda_\text{exc} \approx 43.0\ \mathrm{cm}\), which is comparable to the radial size of the reactor (\(R= 38\ \text{cm}\)). 
At this point, the reflected wave from the reactor wall and the incident wave superimpose on each other and form a standing wave. 
Then the strongest (smallest) electric field is at the antinode (node), resulting in obvious radial nonuniformity of the electron density.
Most importantly, the profiles of the electron density distributions in cases (000) and (111) are the same (see Fig.~\hyperlink{fig2}{2(a)}), which explicitly confirms the similarity of the electromagnetic wave-driven rf plasma under designed different dimensional scales.

By comparing discharge characteristics varied along the pathway \((000)\rightarrow ... \rightarrow (111)\), the effects of \([p,g,f]\) on the discharge uniformity can be comprehensively understood. 
From \((000)\rightarrow (100)\), \((010)\rightarrow (110)\), and \((001)\rightarrow (101)\), one can conclude that increasing the gas pressure has a minor effect on discharge uniformity.
Reducing the geometric dimensions, e.g., \((000)\rightarrow (010)\), can significantly improve the plasma uniformity (mainly related to the reduction of electrode radius), whereas increasing the driving frequency, e.g., \((000)\rightarrow (001)\), substantially exacerbates the radial nonuniformity. 
From \((000)\rightarrow(001)\) and \((100)\rightarrow(101)\), with the driving frequency doubled, we observed a stop band \coloredcite{Lee_2008} in the electron density distribution \coloredcite{Liu2021PST} where the electron density is highly damped from the discharge edge to the center; for example,
in case (001) the electron density is the smallest in the region of \(-0.25 < r/R < 0.25\).
At \(r/R=0.31\), the electron density rapidly increases in the \(r-\)direction and reaches the peak at \(r/R=0.67\). 
Under this condition, the calculated plasma excitation wavelength is \(\lambda_\text{exc} \approx 17.4\ \mathrm{cm}\), much less than the electrode radius \(R_1 = 38\ \mathrm{cm}\).
Figure~\hyperlink{fig2}{2(b)} demonstrate the scale-invariant radial distributions of electron density \(n_e\) under similar discharge conditions, that is \(n_e\) in case (111) is four times that of (000) with \(k = 2\), i.e., \(n_e^{(111)}=4n_e^{(000)}\) or equivalently \(n_e^{(000)} = k^{-2} n_e^{(111)}\). 
From the transformation in Eq.~\coloredeqref{eq-G1Gk}, the present result explicitly confirms the similarity factor for electron density \(\alpha[n_e]=-2\), which extends the application of the similarity law to electromagnetic regimes.

\begin{figure}[htbp]
\hypertarget{fig3}{}
\centering
\includegraphics[clip, width=\linewidth]{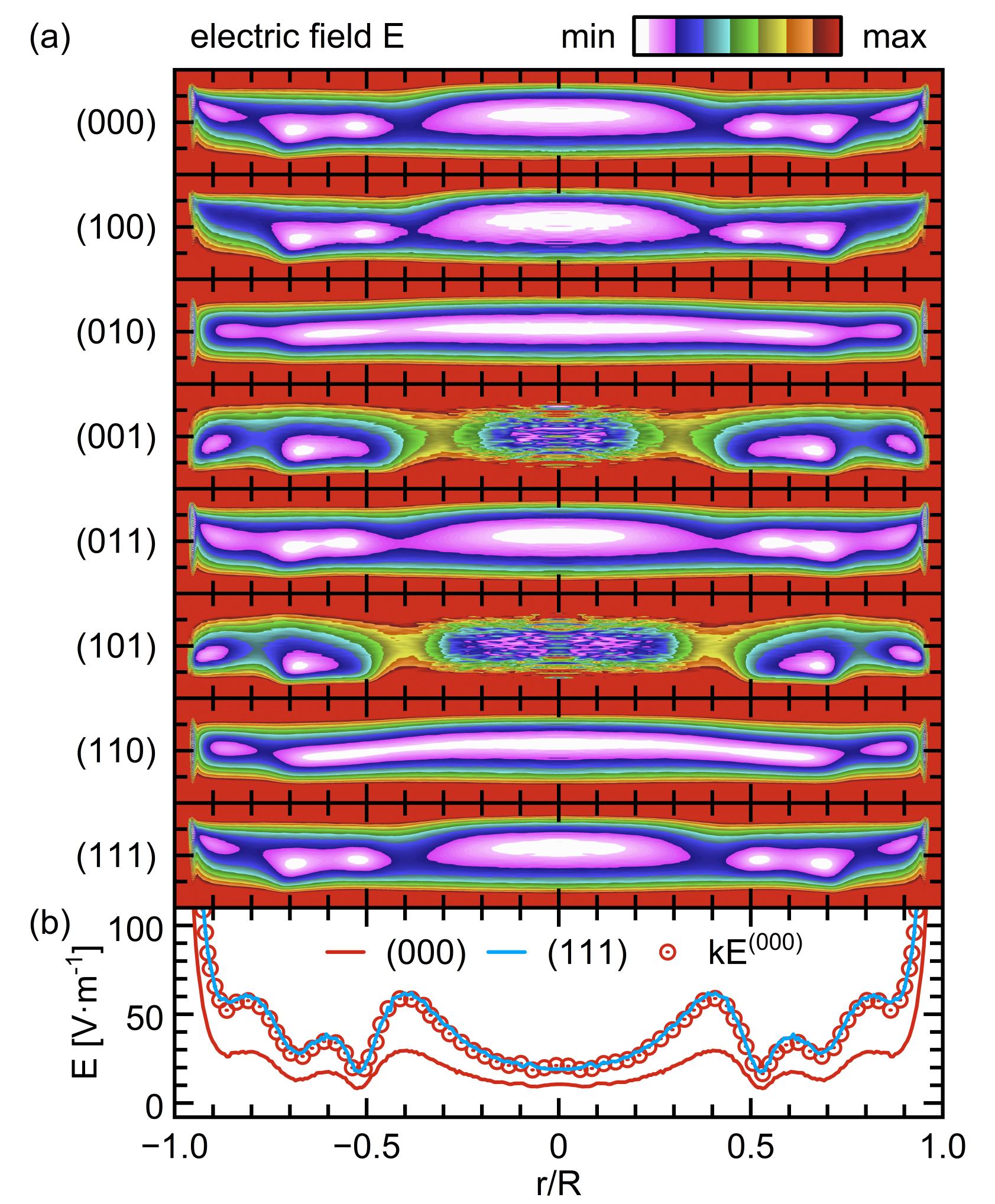}
\caption{\label{fig:Fig3} 
(a) Distributions of the normalized electric field for the base case (000), intermediate cases (100), (010), (001), (011), (101) and (110), and similarity case (111). (b) Radial distributions of the electric field at \(z=d/2\).
The scaled electric field distributions for the base case and similarity case overlap.}
\end{figure}

Figure~\hyperlink{fig3}{3} shows the spatial distributions of the normalized electric field strength for two similar rf discharges, case (000) and (111), and six intermediate states.
The electric field distributions exhibit distinct characteristics under varied conditions, which correspond to the electron density distributions in Fig.~\hyperlink{fig2}{2}.
The electric field strength is calculated from the PIC results via \(E=\sqrt{E^2_r+E^2_z}\), where \( E_r\) and \( E_z \) are the radial and axial components, respectively.
The electric field (primarily axial component) is the strongest in the sheath region, whereas the radial component can be observed across the bulk region. 
The radial distribution of the electric field can be qualitatively described by the uniform slab model \coloredcite{Chabert-2005prl} with the plasma nonuniformity induced by the standing wave effects.

In the base case (000), the electric field is the smallest at the gap center, showing multiple nodes in the radial direction. 
According to the uniform model \coloredcite{Lieberman2002PSST}, \( E_r \) is zero at the center, and the theoretical position of its first extreme point is at \(r=\chi'_{11}/(2\pi/\lambda_\text{exc}) \approx 12.6\ \text{cm}\) (agrees with the PIC result \(r = 14.82\ \text{cm}\)), where \(\chi'_{11}\) is the first zero point of the derivative of the first-order Bessel function.
For the pathway \((000)\rightarrow ... \rightarrow (111)\), the variation of the electric field is consistent with the uniformity of the electron density distributions in Fig.~\hyperlink{fig2}{2(a)}.
Taking the case (010) as an example, the excitation wavelength is \(\lambda _{\text{exc}} \approx 63.01\ \text{cm}\), corresponding to \( r=\chi_{01}/(2\pi /\lambda_\text{exc})=24.12\ \text{cm} > 19\ \text{cm}\), where \(\chi_{01} = 2.405\) is the first zero point of the zero-order Bessel function. The standing wave can not form since the electrode radius is much smaller than the excitation wavelength. The electric field distribution profiles are the same for cases (000) and (111). As shown in Fig.~\hyperlink{fig3}{3(b)}, the scaled radial electric field distributions at \(z=d/2\) for cases (000) and (111) are overlapping, which confirms the similarity factor \(\alpha[\mathbf{E}]=-1\) for the electric field in similar discharges.

\begin{figure}[htbp]
\hypertarget{fig4}{}
\centering
\includegraphics[clip, width=\linewidth]{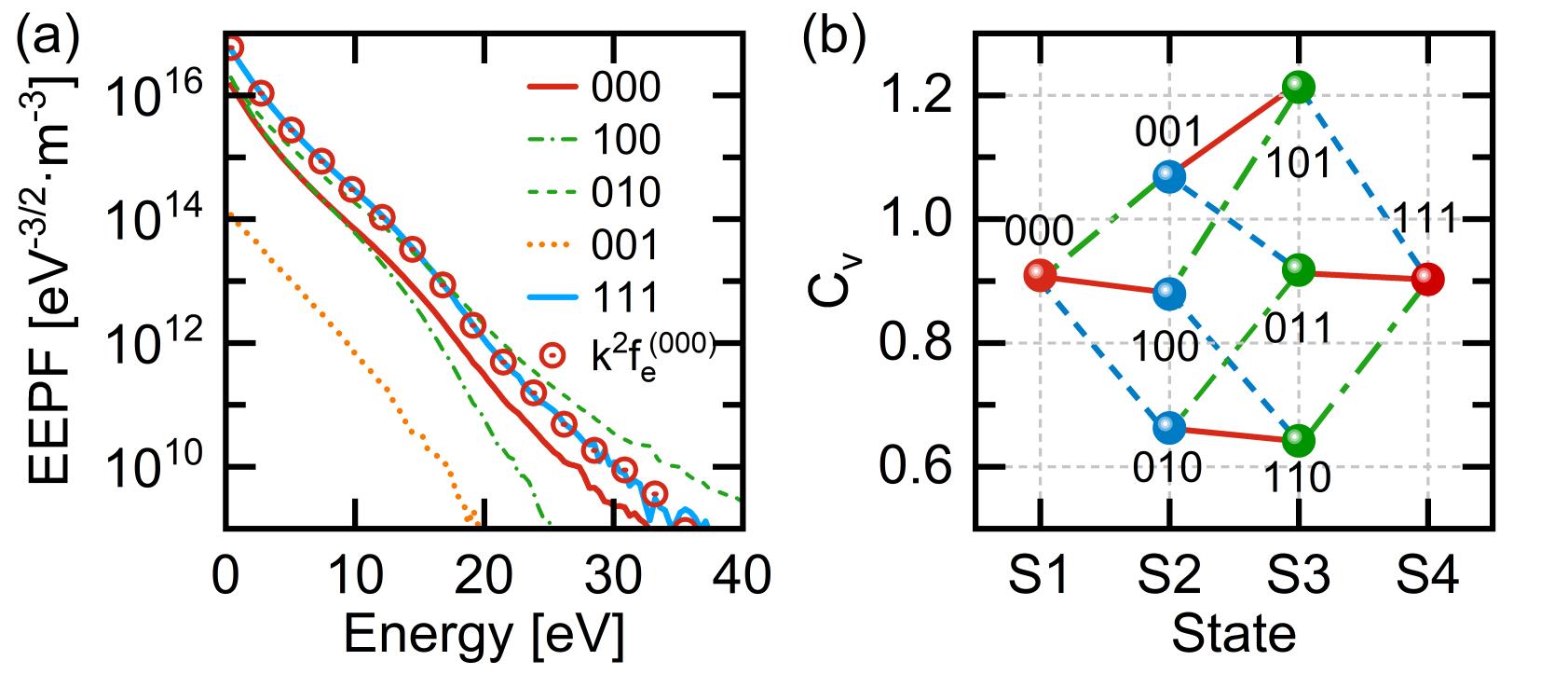}
\caption{\label{fig:Fig4}
(a) Electron energy probability functions for compared cases. (b) Scaling networks for the plasma nonuniformity parameter \(C_\text{v}\) with \([p\), \(g\), \(f]\) gradually tuned at different states (S1--S2--S3--S4).}
\end{figure}

Figure~\hyperlink{fig4}{4(a)} shows the electron energy probability functions (EEPFs) with the control parameters \([p,g,f]\) individually tuned. From \((000)\rightarrow(100)\), the increase in gas pressure shortens the electron mean free path \(\lambda_e\), reducing the energy gain of electrons between two consecutive collisions, and thus leads to the decrease in the population of high-energy electrons, showing a suppressed high-energy tail in EEPF.
For the transition \((000)\rightarrow(010)\), the gap distance is halved, and the mean electric field strength is intensified; meanwhile, the Knudsen number \(\lambda_e/d\) increases, enabling electrons to accumulate more energy during their transit across the gap, thus obviously enhancing the number of medium-to-high energy electrons.
When the driving frequency \(f\) increases, \((000)\rightarrow(001)\), the EEPF curve shifts distinctly downward across the entire energy spectrum. This phenomenon is attributed to the significant reduction in the electron density at the gap center with the presence of the stop band   under higher frequency conditions (see case (001) in Fig.~\hyperlink{fig2}{2(a)}).
Notably, even in the presence of significant electromagnetic effects, the scaled EEPF of (000), i.e., \(k^2f_e^{(000)}\), overlaps with case (111), which confirms the scale-invariance of electron kinetics in similar discharges (see the derivation from Eq.~\eqref{eq:boltzmann02}). The scale-invariant distributions of plasma species establish the foundation of discharge similarity under varied conditions.

Figure~\hyperlink{fig4}{4}(b) shows the scaling networks of the parameter for describing the plasma uniformity at varied states (S1--S4). 
The coefficient of variation \(C_\text{v}\), representing the uniformity of electron density, is expressed as 
\begin{equation}
    C_\text{v}=\frac{n_\text{st}}{\langle n_{{e}} \rangle} =\frac{\sqrt{S^{-1}\int\left( n_{e} - \langle n_{\text{e}} \rangle \right)^2 \text{d}S}}{ S^{-1} \int n_{{e}} \text{d}S},
\end{equation}
where \(S\) is the space area, \(\langle n_{{e}} \rangle\) is the spatially averaged electron density, and \(n_\text{st}\)
is the standard deviation of the electron density.
For \((000)\rightarrow(010)\),
\((100)\rightarrow(110)\), \((001)\rightarrow(011)\), and \((101)\rightarrow(111)\), the halved gap dimensions result in better uniformity. 
For \((000)\rightarrow(001)\), \((100)\rightarrow(101)\), \((010)\rightarrow(011)\), and \((110)\rightarrow(111)\), the doubled frequency exacerbates the plasma uniformity. 
The state S4 has the same value of \(C_\text{v}\) as S1, illustrating how the combined parameter can impact the discharge nonuniformity, which explicitly demonstrates the diverse scaling pathways in a closed loop.
As the discharge state S1 is gradually adjusted toward to similarity state S4, the similarity-based scaling network is formed, in which the same color-coded lines denote the same parameter scaling. The similarity-based scaling network unveils the intricate relationship between control parameter variations, such as the electron density and density nonuniformity.

\textit{Conclusions}---In conclusion, in this Letter we have established the similarity and scaling laws for electromagnetic wave-driven rf discharges.
The scale-invariant distributions of electron density, electric field, and EEPF in geometrically similar systems are demonstrated via 2D electromagnetic PIC simulations, which successfully extend the similarity theory from the electrostatic to electromagnetic regimes.
By gradually tuning the control parameters (gas pressure, gap dimension, and driving frequency), the scaling networks are established, which can effectively correlate plasma states and thus provide reliable prediction, regulation, and optimization strategies for plasma parameters in different plasma systems.  
It is worth emphasizing that the nonlinear physical processes (e.g., plasma resonance heating \coloredcite{fu-prapplied}) and the ion or electron-induced secondary electron emissions do not violate the discharge similarity, even across the electrostatic-to-electromagnetic regimes. 
The similarity theory satisfactorily holds for low-pressure rare gas plasmas and is even more rigorous for pure electron systems, e.g., electron emission driven diodes \coloredcite{lcb-pre, lcb-edl}; however, the discharge in electronegative molecular gases (e.g., \(\text{CF}_4\) \coloredcite{Yang-2025}) with complex reactions may undermine the discharge similarity with approximations. The application of similarity methods can strongly aid in establishing standard diagnostics, cross-comparison, and optimization for plasma equipment in semiconductor manufacturing.


\textit{Acknowledgments}---This work was supported by the National Natural Science Foundation of China (Grant No.~52250051). D.E. acknowledges funding from the German Research Foundation (Project No.~550860775).

\textit{Data availability}---The data that support the findings of this article are not publicly available. The data are available from the contact author upon reasonable request. 

\bibliography{references}
\end{document}